\documentclass[aps,prl,twocolumn,superscriptaddress,longbibliography,nobibnotes,nodoi,nourl,noeprint]{revtex4-2}
\usepackage{amsmath,amssymb,physics,bm,siunitx}
\usepackage{xcolor,graphicx}
\usepackage[colorlinks=true,citecolor=blue,urlcolor=blue,linkcolor=black]{hyperref}
\usepackage{setspace}

\begin{document}

\title{\textbf{Hidden periodicities allow the prediction of locked particle motions\\
 on quasicrystalline surfaces}} 

\author{Seemant Mishra} 
\email{semishra@uos.de}
\affiliation{Universit\"{a}t Osnabr\"{u}ck, Institut für Physik, Barbarastra{\ss}e 7, D-49076 Osnabr\"uck, Germany}

\author{Artem Ryabov} 
\email{artem.ryabov@matfyz.cuni.cz} 
\affiliation{Charles University, Faculty of Mathematics and Physics, Department of Macromolecular Physics, V Hole\v{s}ovi\v{c}k\'{a}ch 2, CZ-18000 Praha 8, Czech Republic}

\author{Philipp Maass} 
\email{maass@uos.de}
\affiliation{Universit\"{a}t Osnabr\"{u}ck, Institut für Physik, 
Barbarastra{\ss}e 7, D-49076 Osnabr\"uck, Germany}

\date{June 4, 2026}

\begin{abstract}
Motion of particles across quasicrystalline surfaces exhibits peculiar features due to the presence of long-range order without translational periodicity.  Under time-periodic forcing, this motion can become locked in directions that deviate strongly from the mean driving direction.
We show that for surface potentials with a quasicrystalline pattern of minima generated by a superposition of plane waves, particle trajectories are nonperiodic, yet their mean direction and speed are determined by hidden periodic potentials. The lattice vectors of these underlying potentials define characteristic velocities that dictate both directional and speed locking. 
The particle motion does not synchronize with the driving, and it is possible for the mean speed to remain nonlocked even in directionally locked states. 
These findings are demonstrated using a model directly amenable to experimental realization.
\end{abstract}

\maketitle

Quasicrystals are structures lacking translational symmetry but exhibiting long-range
order that manifests itself in Bragg diffraction patterns~\cite{Shechtman/etal:1984,Levine/Steinhardt:1984, Macia-Barber:2021, Nagaoka/etal:2023}. First
observed in 
metallic alloys \cite{Gratias/Quiquandon:2019,
  Shechtman/Blech:1985,Ishimasa/etal:1985, Bancel/Heiney:1986},
quasicrystalline order has since been identified
in a wide range of systems, 
including
photonic~\cite{Kohmoto/etal:1987, Vardeny/etal:2013,WangP/etal:2024,Steurer/Sutter-Widmer:2007},
acoustic \cite{Steurer/Sutter-Widmer:2007, Han/etal:2025}, naturally
occurring minerals~\cite{Bindi/etal:2009}, liquid crystals \cite{Zeng/etal:2023}, 
cold atom
\cite{Viebahn/etal:2019,Sbroscia/etal:2020}, mesoporous silica \cite{Xiao/etal:2012}, soft
matter~\cite{Zeng/etal:2004,Fischer/etal:2011,Gillard/etal:2016}, and thin films \cite{Foerster/etal:2013, Haller/etal:2026}. 
Quasicrystals arise from different growth modes
\cite{Schmiedeberg/Stark:2012, Achim/etal:2014, Martinsons/Schmiedeberg:2018, WangL/etal:2020} 
and their
aperiodic yet ordered arrangement gives rise to unusual properties,
such as forbidden rotational
symmetries~\cite{Levine/Steinhardt:1984,Wang/etal:1987},
phasons~\cite{Jiang/etal:1992,Steinhardt/Ostlund:1987,Kromer/etal:2012, Zheng/etal:2025}, exotic
magnetic orders \cite{Tamura/etal:2021},
anomalous thermal~\cite{Archambault/Janot:1997, Nagai/etal:2024} and
electrical~\cite{Pierce/etal:1994} conductivity.
Surfaces of quasicrystals  are known to exhibit
low friction and high oxidation resistance \cite{Thiel:2008, Rabson:2012}. 

A challenging problem in this field is understanding the mechanisms of atomic motion and self-assembly on quasicrystalline surfaces \cite{Franke/etal:2002, Sharma/etal:2008, Smerdon/etal:2008, Smerdon:2010, Im/etal:2026}. Microscopic insights into 
these dynamics are provided by simulations \cite{Kromer/etal:2013, Schoberth/etal:2016, Manini/etal:2025}, and by 
experiments tracking colloidal particles on patterned surfaces \cite{Su/etal:2017}
and within optical 
lattices \cite{Mikhael/etal:2008, Mikhael/etal:2011, Bohlein/Bechinger:2012, Mikhael/etal:2010, Zaidouny/etal:2014}.
Such studies have shown that colloidal monolayers can organize into Archimedean-like tilings. 
Simulations of driven interacting particles on an array with fivefold
rotational symmetry revealed 
the emergence of diverse ordered nonequilibrium states \cite{Reichhardt/OlsonReichhardt:2011, OlsonReichhardt/Reichardt:2012}.
In states with Archimedean tiling, the center-of-mass
velocity can get locked, 
whereby the direction and speed of motion remain invariant despite
slight variations in the angle of the driving force.

The observation of directional locking on a quasicrystalline lattice raises the questions
of whether general principles exist to theoretically predict locked directions of particle motion
and whether directional locking is accompanied by a locking of the particle speed. 

Here we consider 
potentials with a quasicrystalline pattern of minima 
formed by a superposition of (nonpropagating) plane waves of equal
shape in different directions with $n$-fold rotational symmetry ($n
\ne 2, 3, 4, 6$),
\begin{gather}
U(\bm r) = \sum_{j=1}^n u(\bm k_j \cdot \bm r)\,,\\
\bm k_j=\frac{2\pi}{\lambda}\left[\cos(\frac{2\pi(j\!-\!1)}{n})\;\hat{\bm{x}}
+\sin(\frac{2\pi(j\!-\!1)}{n})\;\hat{\bm{y}}\right]
\label{eq:U}
\end{gather}
For driven particle motion on such quasicrystalline surfaces, 
we state the following conjecture, which arises from considering hidden translational periodicities:\\[0.5ex]
\emph{Particle velocities are dictated by symmetries of the periodic $(ij)$-potentials
\begin{equation}
u^{(ij)}(\bm r) = u(\bm k_i \cdot \bm r) + u(\bm k_j \cdot \bm r), \quad i \ne j,
\label{eq:P_potential}
\end{equation}
formed by a superposition of two of the plane waves.}

\vspace{0.5ex}
The minima of each $(ij)$-potential form a periodic lattice spanned by 
primitive vectors $\bm{a}^{(ij)}_\alpha$, $\alpha = 1, 2$, satisfying 
\begin{equation}
\bm{a}^{(ij)}_\alpha \cdot \bm{k}^{(ij)}_\beta  = 2\pi \delta_{\alpha\beta}, 
\label{eq:a-vectors}
\end{equation}
where $\bm{k}_1^{(ij)} = \bm{k}_i$ and $\bm{k}_2^{(ij)} = \bm{k}_j$.

\begin{figure}
\includegraphics[width=\linewidth]{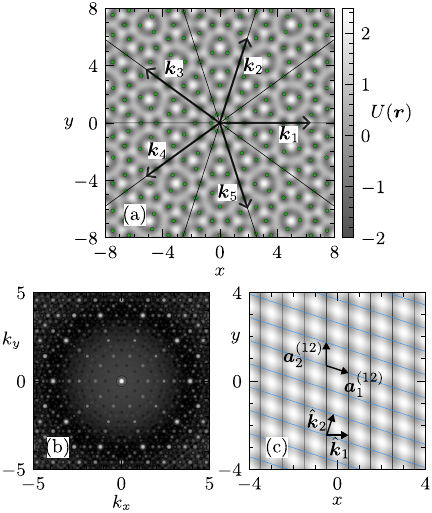}
\caption{(a) Ten-fold rotationally symmetric potential $U(\bm r)$
generated by five sinusoidal
plane waves with wave vectors
$\bm k_j$ [Eqs.~\eqref{eq:U}, \eqref{eq:QPP_form}]. Its minima are indicated by green
bullets.
(b) Bragg diffraction pattern of the
quasicrystalline lattice of potential minima. (c) Hidden periodic potential $u^{(12)}(\bm r)$ [Eq.~\eqref{eq:P_potential}]. Minima of $u^{(12)}(\bm r)$ form an oblique lattice spanned
by primitive vectors $\bm a^{(12)}_1$ 
and $\bm a^{(12)}_2$ orthogal to the wavevectors $\bm k_2$ and $\bm k_1$, respectively
[Eq.~\eqref{eq:a-vectors}].}
\label{fig:Upotential}
\end{figure}

Figure~\ref{fig:Upotential}(a) shows an example of $U(\bm r)$ for $n=5$ sinusoidal
plane waves.
Its minima form a ten-fold symmetric quasicrystalline lattice whose 
Bragg diffraction pattern is given in Fig.~\ref{fig:Upotential}(b). 
Figure~\ref{fig:Upotential}(c) illustrates the hidden periodic $(12)$-potential.

For particles driven by a time-periodic force 
\begin{equation}
\bm{F}(t) = \bm{F}_{\rm dc} + \bm{F}_{\rm ac}(t),
\end{equation}
with $\bm{F}_{\rm ac}(t+\tau) = \bm{F}_{\rm ac}(t)$ a $\tau$-periodic
function having zero mean, the conjecture manifests itself in the following 
properties:
\newcounter{listnumber}
\begin{list}{\arabic{listnumber}.}{\setlength{\leftmargin}{1.1em}\setlength{\rightmargin}{0em}
\setlength{\itemsep}{0ex}\setlength{\topsep}{0.5ex}\setcounter{listnumber}{1}\usecounter{listnumber}}

\item The mean particle velocities $\bar{\bm{v}}$ can
lock into symmetry directions determined by lattice vectors 
\begin{equation}
\bm{d}^{(ij)}_{m_1m_2} = m_1\bm{a}^{(ij)}_1 + m_2\bm{a}^{(ij)}_2,
\label{eq:d_vector}
\end{equation}
of the $(ij)$-potentials, where $m_1, m_2 \in \mathbb{Z}$ are coprime. In other words,
$\bar{\bm{v}}$ points in the same direction as $\bm{d}^{(ij)}_{m_1m_2}$
within certain regions of the space of driving parameters. 
This direction can deviate from that of $\bm{F}_{\rm dc}$.
    
\item The mean particle speed $\bar{v} = |\bar{\bm{v}}|$ can lock
  into rational multiples
\begin{equation}
\bar{v} = \frac{p}{q}\, {v}^{(ij)}_{m_1m_2}, \quad p,q \in \mathbb{N},
\label{eq:barv-vcsingle}
\end{equation}    
of the characteristic speed 
\begin{equation}
v^{(ij)}_{m_1m_2} = \frac{|\bm{d}^{(ij)}_{m_1m_2}|}{\tau}\,.
\label{eq:vc-single}
\end{equation} 
Frequently, direction and speed locking occur
simultaneously, implying $\bar{\bm{v}} = (p/q)
(\bm{d}^{(ij)}_{m_1m_2}/\tau)$. However, contrary to particle motion 
in periodic potentials, dynamics are not synchronized and 
particle trajectories are not periodic.

\item It can happen that lattice vectors $\bm{d}^{(ij)}_{m_1m_2}$ and 
$\bm{d}^{(kl)}_{m_1'm_2'}$ 
of different  $(ij)$- and $(kl)$-potentials point in the same direction,
$\bm{d}^{(ij)}_{m_1m_2} \parallel \bm{d}^{(kl)}_{m_1'm_2'}$, but have different
lengths, $|\bm{d}^{(ij)}_{m_1m_2}|\ne|\bm{d}^{(kl)}_{m_1'm_2'}|$.
In that case it is possible that the particle's mean speed also locks
into $\bar v=(p/q)v^{(ij,kl)}_{\rm c}$
with a characteristic
velocity
\begin{equation}
v^{(ij,kl)}_{\rm c} = 
\frac{n_1|\bm{d}^{(ij)}_{m_1m_2}|\!+\!n_2|\bm{d}^{(kl)}_{m_1'm_2'}|}{\tau}\,,
\label{eq:vc-mixed}
\end{equation}
where $n_1,n_2\in\mathbb{Z}\backslash \{0\}$ are coprime.

\end{list}

\begin{figure*}
\includegraphics[width=\textwidth]{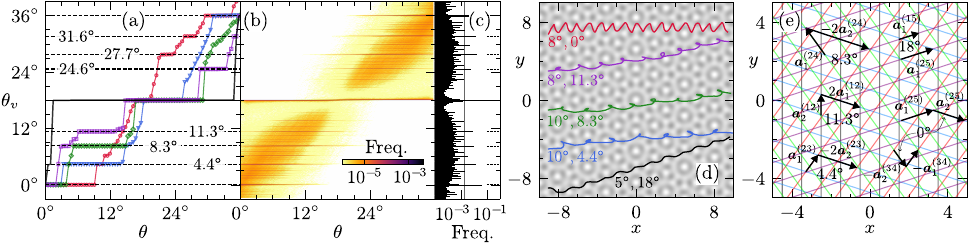}
\caption{Directional locking of particle motion when driven by the force $\bm{F}_{\rm{dc}} + \bm{F}_{\rm{ac}}(t)$ [Eqs.~\eqref{eq:Fdc}, \eqref{eq:Fac}]
across the potential $U(\bm r)$ of Fig.~\ref{fig:Upotential}(a).
In (a)-(d), $\bm F_{\rm dc}=10\hspace{0.1em}(\cos\theta,\sin\theta)$.
(a) Angle $\theta_v$ between mean particle velocity $\bar{\bm{v}}$ and $x$-axis
as a function of the mean driving angle $\theta$ for $(F^{\rm ac}_x,F^{\rm ac}_y)=(0,0)$
(black line), and for ac driving with
$(F^{\rm ac}_x,F^{\rm ac}_y)=(20,7)$ and $1/\tau=3.9$ (blue triangles), 4.4 (green diamonds), and
4.9 (purple squares). The red circles are for $(F^{\rm ac}_x,F^{\rm ac}_y)=(0,20)$ and
 $1/\tau=7$. Dashed horizontal lines indicate robust locking directions. 
(b)  Color plot with logarithmic scale of relative counts (frequencies) of $\theta_v$ 
for $\theta\in[0^\circ,36^\circ)$. For each $\theta$ we have generated
$5\times 10^4$ trajectories for randomly chosen $1/\tau\in [2,12]$
and $(F^{\rm ac}_x,F^{\rm ac}_y)$ satisfying $[(F^{\rm ac}_x)^2+(F^{\rm ac}_y)^2]^{1/2}\le30$. Resolution of the data sampling
 corresponds to $10^3$ and 200 bins along the 
$\theta_v$- and $\theta$-axes.
(c) Log-scale histogram of $\theta_v$ obtained by
summing over $\theta$ in (b). Spikes occur for robust locking directions.
The vertical line indicates the frequency $1.88\times10^{-2}$
above which all locking directions follow from combining primitive lattice vectors of 
a single periodic $(ij)$-potential.
(d) Particle trajectories for locked directions in (a), where $\theta$, $\theta_v$-values
are given on the left.
(e) Integer linear combinations of primitive lattice vectors
yielding locked directions in (a). The complete set of linear combinations is given
in supplemental material (see below).}
\label{fig:directional-locking}
\end{figure*}

In principle, with $(m_1,m_2) \in \mathbb{Z}^2$, pointing of lattice vectors
$\bm{d}^{(ij)}_{m_1m_2}$ can be arbitrarily close to any direction. However,
as discussed below, well observable robust locking in extended 
parameter regions occurs for small $m_1$ and $m_2$. The same holds
true for values of $p$, $q$, and $n_1$, $n_2$. 

\vspace{0.5ex}
\textit{Model for testing conjecture.}
We now show how the conjecture and its implications can be applied to identify locked states of driven particle 
motion on a quasicrystalline surface. To this end,
we consider the quasiperiodic
potential~\eqref{eq:U} with $n=5$ and 
\begin{equation} 
u(\bm{k}_j \cdot \bm{r})=\frac{u_0}{2} \cos(\bm{k}_j \cdot \bm{r})\,,
\label{eq:QPP_form}
\end{equation}
corresponding to the optical potential used in experiments on overdamped dynamics of colloidal particles
\cite{Bohlein/Bechinger:2012}. The particle dynamics follows the Langevin equation
\begin{equation}
\gamma\dot{\bm{r}} = -\bm{\nabla} U(\bm{r}) + \bm{F}_{\rm{dc}} + \bm{F}_{\rm{ac}}(t) + \sqrt{2D}\,\boldsymbol{\eta}(t)\,, 
\label{eq:langevin}
\end{equation}
where the friction coefficient $\gamma$ and the temperature
$T$ give the diffusion coefficient $D= k_{\rm B} T /\gamma$, and $\bm{\eta}(t)$ is a Gaussian white noise process with $\langle \eta_\alpha(t) \rangle=0$ and $\langle \eta_\alpha(t)\eta_\beta(t') \rangle = \delta_{\alpha\beta}\delta(t-t')$.  For the driving, we use 
\begin{align}
\bm{F}_{\rm dc} &= F_{\rm dc} \left( \cos\theta, \sin\theta  \right), 
\label{eq:Fdc}\\
\bm{F}_{\rm ac}(t) &= \left( F^{\rm ac}_x \cos(2 \pi t/\tau), F^{\rm ac}_y \sin(2 \pi t/\tau)  \right)\,.
\label{eq:Fac}
\end{align}  
The set of all possible values of $F_{\rm dc}, \theta, F^{\rm ac}_x, F^{\rm ac}_y$, and the frequency $1/\tau$ defines the space of driving parameters. As units of length, time, and force we take $\lambda, \lambda^2 \gamma/u_0$ and $u_0/\lambda$, respectively. Due to the ten-fold symmetry of the quasiperiodic potential, we restrict our 
analysis to $\theta\in[0,36^\circ)$.

For the sake of clarity, we present our results for  $D=0$ and initial conditions $x(0)=y(0)=0$. 
The impact of noise or of changing initial conditions is to decrease or modify the robustness of locking but not to alter
possible locked directions and speeds. This is  
discussed in supplemental material, see below. 

\vspace{0.5ex}
\textit{Directional locking.}
For quantifying directional locking, we introduce the angle
\begin{equation}
\theta_v=\arctan\frac{\bar v_y}{\bar v_x}\,,
\label{eq:thetav}
\end{equation}
between the $x$-axis and mean particle velocity $\bar v=(\bar v_x,\bar v_y)$ in the steady state.
The motion is directionally locked if $\theta_v$ does not change
upon variation of the dc-driving direction specified by $\theta$.
For $\bm{F}_{\rm ac}=\bm0$, $\theta_v$ is primarily locked
to the direction of $18^\circ$ for an extended interval of $\theta$ values.
Only for very small $\theta$-intervals, $\theta_v$ locks into
other directions. An example is shown in Fig.~\ref{fig:directional-locking}(a) by the black line.

Pronounced additional locking directions become visible for $\bm{F}_{\rm ac}\ne\bm0$, 
where locked $\theta_v$ different from $18^\circ$ appear 
as broad plateaus in the step-like behavior of the $\theta_v$-$\theta$ relation. 
Figure~\ref{fig:directional-locking}(a) shows a few representative examples. 

To evaluate the statistical significance of the locking directions, we 
have chosen 200 equally spaced $\theta\in [0^\circ, 36^\circ)$. For each $\theta$,
we have performed
$5\times 10^4$ simulations for fixed $F_{\rm dc}=10$, and randomly chosen
$[(F_x^{\rm ac})^2+(F_y^{\rm ac})^2]^{1/2}\le 30$ and $1/\tau\in [2, 12]$.
Moreover, the ellipses of the ac-forcing in Eq.~\eqref{eq:Fac} were randomly rotated.
Frequencies of $\theta_v$ occurring at a driving direction $\theta$
are shown in the color plot of Fig.~\ref{fig:directional-locking}(b).
Most significant locking directions appear as extended horizontal dark lines in this plot:
apart from the one for $18^\circ$, there are many extended lines
for other directions, including those identified in
Fig.~\ref{fig:directional-locking}(a). The dark lines correspond to delta-like peaks in the
histogram of $\theta_v$-values shown in Fig.~\ref{fig:directional-locking}(c)
for the sampled parameter space.

The directional locking does not imply that the particle trajectories are periodic as it would be for dynamics synchronized with the ac-driving in periodic potentials. This is demonstrated in Fig.~\ref{fig:directional-locking}(d), where we show the trajectories in locked states corresponding to the broadest plateaus in Fig.~\ref{fig:directional-locking}(a).

The locking directions can be explained based on the hidden periodicities of $U(\bm r)$ by a simple geometric construction: for each
of the five plane waves $\cos(\bm k_j\cdot\bm r)$,
we draw the isolines of their minima.
This gives the straight lines in Fig.~\ref{fig:directional-locking}(e), colored according to the different $\bm k_j$.
Two sets of identically colored lines for $\bm k_i$, $\bm k_j$ intersect at the $(ij)$-sublattice formed by
the minima of the periodic $(ij)$-potential.
The primitive lattice vectors $\bm a_1^{(ij)}$, $\bm a_2^{(ij)}$ 
from Eq.~\eqref{eq:a-vectors} connect nearest-neighbors
of the $(ij)$-sublattice, see also Fig.~\ref{fig:Upotential}(c).

The lattice vectors $\bm{d}^{(ij)}_{m_1m_2} = m_1\bm{a}^{(ij)}_1 + m_2\bm{a}^{(ij)}_2$
in Eq.~\eqref{eq:d_vector} give possible directions of locking. For the
locked values 
$\theta_v\cong 0^\circ, 4.4^\circ, 8.3^\circ, 11.3^\circ, 18^\circ$
in Fig.~\ref{fig:directional-locking}(a), the construction of corresponding
$\bm{d}^{(ij)}_{m_1m_2}$ is illustrated in Fig.~\ref{fig:directional-locking}(e).
It is possible that multiple lattice vectors
give the same $\theta_v$, see, for example,
the $\bm d_{11}^{(25)}$, $\bm d_{-1\,-1}^{(34)}$ for
$\theta_v=0^\circ$, and the $\bm d_{10}^{(15)}$, $\bm d_{10}^{(25)}$
for $18^\circ$. We call such angles $\theta_v$ degenerate.
The 11 most frequently occurring $\theta_v$ in Fig.~\ref{fig:directional-locking}(c) all 
correspond to certain $\bm{d}^{(ij)}_{m_1m_2}$ with small $|m_j|$: $|m_j|\le2$ for the
most frequent 10 ones, and $(m_1,m_2)=(-3,1)$ in one case.

Differently speaking, all peaks in
Fig.~\ref{fig:directional-locking}(c) passing the vertical solid line at threshold 
frequency $1.88\times 10^{-3}$
correspond to directions determined by small-integer linear combinations
of the two lattice vectors of a single $(ij)$-potential.
We refer to these directions as the most robust locking directions.
A table of the corresponding lattice vectors is given in SM. 

\begin{figure}[t!]
\includegraphics[width=\linewidth]{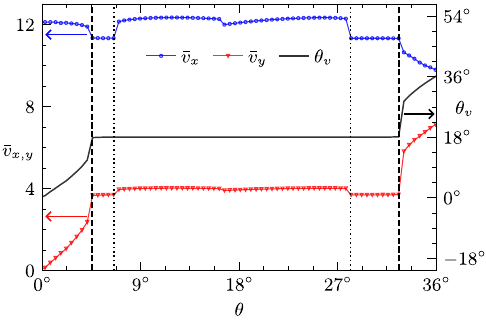}
\caption{Directional locking without  and with speed locking for 
$F_{\rm dc}=14$, $(F^{\rm ac}_x,F^{\rm ac}_y)=(2,2)$, and $1/\tau=7$.
For the $\theta$-interval between the vertical dotted lines,
the mean velocity $(\bar v_x,\bar v_y)$ is varying, 
while $\theta_v=\arctan(\bar v_y/\bar v_x)$ stays constant.
Between the vertical dashed and dotted lines, speed locking 
goes along with directional locking.}
\label{fig:speed_locking_demonstration}
\end{figure}

\vspace{0.5ex}
\textit{Speed locking.}
Since particle dynamics are not synchronized with the 
time-periodic driving and particle trajectories are not periodic, one may wonder
whether the particle's time-averaged speed $\bar v$ can become locked.
Our simulations have shown that speed locking can occur if directional locking is present.
In directionally locked states, the ratio $\bar v_y/\bar v_x$ must be constant
according to Eq.~\eqref{eq:thetav}, but this does not imply that
$\bar v_x$ and $\bar v_y$ and hence $\bar v$ are locked.

\begin{figure}[t!]
\includegraphics[width=\linewidth]{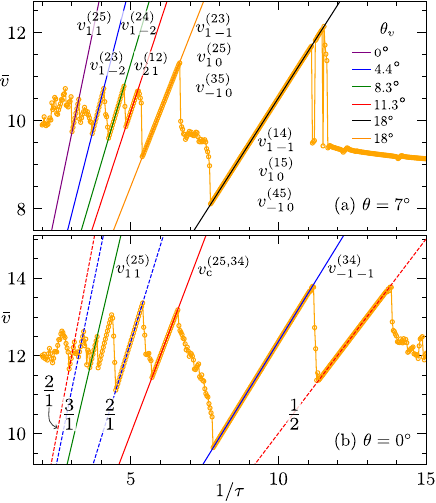}
\caption{Mean particle speed $\bar v$ as a function of the ac-driving frequency $1/\tau$
for (a)
mean driving angle $\theta=7^\circ$ and force parameters $F_{\rm dc}=10$, $(F^{\rm ac}_x,F^{\rm ac}_y)=(20,7)$
as in Fig.~\ref{fig:directional-locking}(a), and (b)
$\theta=0^\circ$, $F_{\rm dc}=12$, and $(F^{\rm ac}_x,F^{\rm ac}_y)=(20,20)$. Circles represent results
from simulations and lines have slopes theoretically predicted when 
applying the conjecture of hidden periodicities. For the labeling of the lines,
see Eqs.~\eqref{eq:barv-vcsingle}-\eqref{eq:vc-mixed}.} 
\label{fig:speed_locking_analysis}
\end{figure}

An example is shown in Fig.~\ref{fig:speed_locking_demonstration},
where $\theta_v$ is locked to $18^\circ$ in the interval 
$4.5^\circ\lesssim\theta\lesssim32^\circ$ bounded by the dashed vertical lines. 
In boundary regions of this interval
between dashed and dotted lines, $\bar v_x$ and $\bar v_y$ are individually
locked. However, this is not the case in the $\theta$-interval between the dotted lines.
In this regime,  $\bar v_x$ and $\bar v_y$ vary with $\theta$ even though
$\bar v_y/\bar v_x$ is constant.

For the most robust locking directions, our sampling of the parameter space
revealed speed locking in 64.5\% of the cases.

Equations~\eqref{eq:barv-vcsingle}, \eqref{eq:vc-single} imply that
$\bar v$ is proportional to the driving frequency $1/\tau$ for speed-locked states.
This dependence is demonstrated in Fig.~\ref{fig:speed_locking_analysis}(a)
for the same force parameters as 
in Fig.~\ref{fig:directional-locking}(a) and driving direction $\theta=7^\circ$:
as a function of $1/\tau$, several regimes appear where 
 $\bar v=|\bm d^{(ij)}_{m_1m_2}|/\tau$.
 That is, the length of the lattice vector $\bm d^{(ij)}_{m_1m_2}$ in the locked direction
 gives the slope of the straight lines in Fig.~\ref{fig:speed_locking_analysis}(a).
 
 In case where locking directions $\theta_v$ are degenerate,
 $\bm d^{(ij)}_{m_1m_2}$ is not unique. Among the 11 most robust locking directions, 
multiple  lattice vectors exist only for  $\theta_v=0^\circ$ and $\theta_v=18^\circ$, where
$\theta_v=0^\circ$ is two-fold and $\theta_v=18^\circ$ is six-fold degenerate.
These degeneracies could be the reason why $\theta_v=18^\circ$ and $\theta_v=0^\circ$ are the most and
second most frequent locking directions. 
  
For $\theta_v=0^\circ$, the two vectors $\bm d^{(25)}_{11}$, $\bm d^{(34)}_{-1\,-1}$ have 
different lengths, $|\bm d^{(25)}_{11}|\cong3.24$ and $|\bm d^{(34)}_{-1\,-1}|\cong1.24$.
The purple line
showing speed locking for $\theta_v=0^\circ$ in Fig.~\ref{fig:speed_locking_analysis}(a) 
has the larger slope. We thus can conclude  that it is the hidden periodic
$(25)$-potential that dictates the average particle motion. 

Among the six vectors $\bm d^{(ij)}_{m_1m_2}$ yielding
$\theta_v=18^\circ$, 
 three have the length $|\bm d^{(14)}_{1\,-1}|=|\bm d^{(15)}_{10}|=|\bm d^{(45)}_{-1\,0}|=1.05$ and the other three
 $|\bm d^{(23)}_{1\,-1}|=|\bm d^{(25)}_{10}|=|\bm d^{(35)}_{-1\,0}|=1.70$. The black and yellow line in 
 Fig.~\ref{fig:speed_locking_analysis}(a) result from speed locking 
 for the smaller and larger slope. We thus gain a reduction of degeneracy: while
 the directional locking alone allows six different periodic $(ij)$-potentials to dictate the 
 average particle motion, only three remain when including the information from the speed locking.

A striking implication of our conjecture are the additional characteristic velocities in Eq.~\eqref{eq:vc-mixed}
that result from two distinct periodic $(ij)$-potentials. The two lattice vectors $\bm d^{(34)}_{-1\,-1}$, $\bm d^{(25)}_{11}$ for $\theta_v=0^\circ$ fulfill the
condition of having the same direction but different lengths.
Therefore, mean particle speed can be locked also according to Eq.~\eqref{eq:vc-mixed}, which is demonstrated in  
Fig.~\ref{fig:speed_locking_analysis}(b):  the solid red line has slope corresponding to
$\bar v=v^{(25,34)}_{\rm c}=(|\bm d^{(25)}_{11}|-|\bm d^{(34)}_{-1\,-1}|)/\tau$ 
[i.e., $n_1=1$, $n_2=-1$ 
in Eq.~\eqref{eq:vc-mixed}]
and we see also
two further speed lockings with $\bar v=2v^{(25,34)}_{\rm c}$ and $\bar v=(1/2)v^{(25,34)}_{\rm c}$
(dashed red lines). In addition, speed locking with $\bar v=p v^{(34)}_{-1\,-1}$, $p=1,2,3$ (blue lines) and
$\bar v=v^{(25)}_{11}$ (green line) occurs.

\vspace{0.5ex}
\textit{Conclusions.} Pronounced
directional and speed locking of particle motion occurs under driving by a dc and ac force across a quasicrystalline surface given by a
superposition of plane waves.
We have proposed a conjecture that predicts
the locked directions and speeds. It is
based on all hidden periodic potentials formed by superimposing two of the plane waves. In spite of the fact that the particle motion is not periodic, the averaged
direction and speed reflect the hidden periodicities.

Our findings raise a number of questions and directions of further research: one 
may ask whether a rigorous proof can be given for the conjecture or at least some of 
its implications. We believe that it is 
possible to observe most robust locking directions given by lattice vectors of the hidden periodic potentials
in experiments similar as in Refs.~\cite{Bohlein/Bechinger:2012, Su/etal:2017}. 
Furthermore, we expect that directional and speed locking in many-particle dynamics
across quasicrystalline surfaces will be governed by
the hidden periodic potentials also.

\vspace{3ex}

\begin{acknowledgments}
We thank the Deutsche Forschungsgemeinschaft (Project No.\ 521001072) and the Czech Science Foundation (Project No.\ 23-09074L) for financial support and gratefully acknowledge the use of a high-performance computing cluster funded by the Deutsche Forschungsgemeinschaft (Project No.\ 456666331).
\end{acknowledgments}



%

\onecolumngrid
\newpage
\renewcommand{\theequation}{S\arabic{equation}}
\renewcommand{\thefigure}{S\arabic{figure}}
\setcounter{equation}{0}
\setcounter{figure}{0}
\begin{center}
\setcounter{page}{7}
{\large\bf Supplemental Material for}\\[2ex]
{\large\bf Hidden periodicities allow the prediction of locked particle motions\\ on quasicrystalline surfaces}\\[2ex]
Seemant Mishra,$^1$ Artem Ryabov,$^2$ and Philipp Maass,$^1$\\[2ex]
$^1$\textit{Universit\"{a}t Osnabr\"{u}ck, Institut für Physik, Barbarastra{\ss}e 7, D-49076 Osnabr\"uck, Germany}\\[1ex]
$^2$\textit{Charles University, Faculty of Mathematics and Physics, Department of Macromolecular Physics,\\ V Hole\v{s}ovi\v{c}k\'{a}ch 2, 
CZ-18000 Praha 8, Czech Republic}
\end{center}

\setstretch{1.5}
\vspace{1ex}\noindent

Section~\ref{sec:minima-of-potential} describes the calculation of the
potential minima shown in Fig.~1(a) and the Bragg peak pattern in
Fig.~1(b).  In Sec.~\ref{sec:dt-ttot-control}, we discuss the choice
of time step for numerical integration and of the total averaging time
for particle velocities. Lattice vectors for the most robust locking
directions are given in
Sec.~\ref{sec:lattice-vectors}. Section~\ref{sec:initial-condition-and-noise}
shows how directional and speed locking depend on initial conditions
and noise.

\section{Minima of the surface potential and Bragg pattern}
\label{sec:minima-of-potential}
For determining the minima of the surface potential given by Eqs.~(1)
and (10) of the main text, we applied the Limited-memory
Broyden-Fletcher-Goldfarb-Shanno algorithm \cite{Liu/Nocedal:1989}. We
performed the search for the minima by starting from initial points on
a square grid of $400 \cross 400$ points within the area
$[-50,50]\times [-50, 50]$ of the $xy$-plane.  In this search, a point
is taken as a candidate of a minimum position if the difference
between an updated and former point is less than $10^{-14}$.  For
multiple minima candidates within a $0.01$ distance, their center of
mass point is taken as the minimum position.  Within the specified
region, a total of $N=8088$ minima were found.

The structure factor is then calculated by
\begin{equation}
S(\bm{k}) = \Biggl| \sum_{j=1}^N w(|\bm r_j|)\exp\left( i \bm{k} \cdot \bm{r}_j \right)\Biggr|^2\,,
\end{equation}
where $\bm r_j$ are the positions of the minima and
\begin{equation}
w(r)=\left\{\begin{array}{c@{\hspace{1em}}c} 
\dfrac{1}{2}\left[1 + \cos\!\left(\dfrac{\pi r}{R}\right)\right], & r \leq R\,, \\
0\,, & r > R\,,
\end{array}\right.
\end{equation}
is the Hann function \cite{Harris:1978} with
$r=\sqrt{x^2+y^2}$ and $R=50$. 
The Hanning window eliminates edge effects.

\section{Control of time step and averaging time used in simulations}
\label{sec:dt-ttot-control}
In the zero-noise limit, the equation of motion (11) of the main text
is numerically integrated by applying the fourth-order Runge-Kutta
method \cite{Press/etal:2007} with a time step $\Delta t$. Particle
positions $\bm r(t)$ are determined for a total time $t_{\rm tot}$ and
the mean velocity is calculated from $\bar{\bm v}\cong\int_0^{t_{\rm
    tot}}\dot{\bm r}(t)\dd t/t_{\rm tot}= \bm r(t_{\rm tot})/t_{\rm
  tot}$ for the initial condition $\bm r(0)=\bm 0$.

\begin{figure}[h!]
\centering
\includegraphics[width=1.0\textwidth]{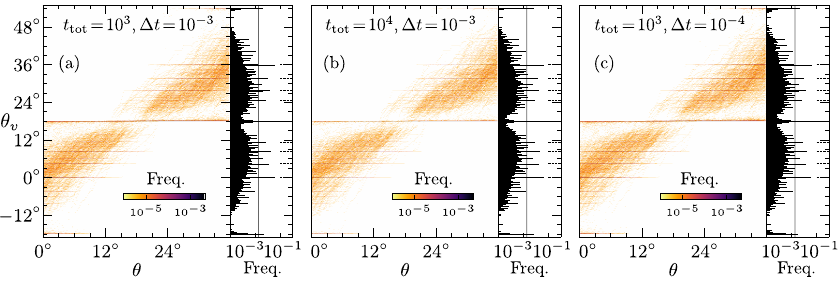}
\caption{Frequencies of velocity angle $\theta_v$ versus driving angle
  $\theta$ as in Figs.~2(b,c) for different time steps $\Delta t$ and
  $t_{\rm tot}$.  The same space of driving parameters as in
  Figs.~2(b,c) is sampled with $10^3$ randomly drawn values for each
  of the 200 equidistant $\theta\in[0^\circ,36^\circ)$.}
\label{fig:dt-ttot-check-histograms}
\end{figure}

In the main text, we present results obtained with $\Delta t=10^{-3}$
and $t_{\rm tot}=10^3$.  We have checked that the time step and
averaging time are sufficiently short and long for obtaining accurate
mean particle velocities, see color plots and histograms in
Fig.~\ref{fig:dt-ttot-check-histograms}.  Decreasing $t_{\rm tot}$ or
increasing $\Delta t$ leads to a slight broadening of the dark lines
in the color plots and accordingly of the peaks in the histograms.

\vspace*{-4ex}
\section{Lattice vectors for most robust locking directions}
\label{sec:lattice-vectors}
Lattice vectors $\bm d^{(ij)}_{m_1m_2}$ of possible locking directions are defined in
Eq.~(6) of the main text.
The 11 most robust locking directions with angle $\theta_v$ in Fig.~2c of the main
text all correspond to directions of lattice vectors of a single $(ij)$-potential. These $\bm
d^{(ij)}_{m_1m_2}$ are listed in
Table~\ref{tab:lattice-vectors}.

For $\theta_v=0^\circ$, there exist two lattice vectors $\bm
d^{(34)}_{-1\,-1}=-\bm a^{(34)}_1-\bm a^{(34)}_2$ and $\bm
d^{(25)}_{1\,1}=\bm a^{(25)}_1+\bm a^{(25)}_2$,
i.e.\ $\theta_v=0^\circ$ is twofold degenerate. The two lattice
vectors $\bm d^{(34)}_{-1\,-1}$ and $\bm d^{(25)}_{11}$ have different
lengths.  This implies the presence of the speeds
$v^{(34)}_{-1\,-1}=|\bm d^{(34)}_{-1\,-1}|/\tau$ and
$v^{(25)}_{11}=|\bm d^{(25)}_{11}|/\tau$ according to Eq.~(8). 
Moreover, we find the further characteristic speed
$v^{(25,34)}_{\rm c}=(|\bm d^{(25)}_{11}|-|\bm d^{(34)}_{-1\,-1}|)/\tau$ in accordance with Eq.~(9).
Speed locking corresponding to these characteristic values is demonstrated
in Fig.~4(b) of the main text.

For $\theta_v=18^\circ$, there are six lattice vectors, where three of
them have length 1.05 and the other three 1.70.

\vspace*{-1ex}
\begin{table}[ht]
\centering
\caption{Angles $\theta_v$ between mean particle velocity
and $x$-axis for the most robust locking directions
and lattice vectors $\bm d^{(ij)}_{m_1m_2}$ 
of a single $(ij)$-potential having the same angle with the $x$-axis.
For $\theta_v=0^\circ$ and $\theta_v=18^\circ$ there exist two and six
different lattice vectors, respectively.}
\label{tab:lattice-vectors}\vspace{1ex}
\resizebox{\textwidth}{!}{
\begin{tabular}{|c||c|c|c|c|c|c|c|c|c|c|c|c|c|c|c|c|c|}
\hline
$\theta_v$ (deg) & 0.00 & 0.00 & 2.98 & 4.39 & 6.18 & 8.27 & 11.35 & 18.00 & 18.00 & 18.00 & 18.00 & 18.00 & 18.00 & 24.65 & 27.73 & 29.82 & 31.61 \\ \hline
$|\bm d^{(ij)}_{m_1m_2}| $          & 1.24 & 3.24 & 3.86 & 2.63 & 4.88 & 2.26 & 2.04 & 1.05 & 1.05 & 1.05 & 1.70 & 1.70 & 1.70 & 2.04 & 2.26 & 4.88 & 2.63 \\ \hline
$i$            & 3    & 2    & 3    & 2    & 2    & 2    & 1    & 1    & 4    & 1    & 2    & 2    & 3    & 3    & 1    & 3    & 2    \\ \hline
$j$            & 4    & 5    & 5    & 3    & 5    & 4    & 2    & 4    & 5    & 5    & 5    & 3    & 5    & 4    & 3    & 5    & 3    \\ \hline
$m_1$          & $-1$ & $1$ & $-3$ & $1$ & $2$ & $1$ & $2$ & $1$ & $-1$ & $1$ & $1$ & $1$ & $-1$ & $-1$ & $2$ & $-2$ & $2$ \\ \hline
$m_2$          & $-1$ & $1$ & $1$ & $-2$ & $1$ & $-2$ & $1$ & $-1$ & $0$   & $0$   & $0$   & $-1$ & $0$   & $-2$ & $-1$ & $-1$ & $-1$ \\ \hline
\end{tabular}
}
\end{table}

\section{Dependence on initial conditions and impact of noise}
\label{sec:initial-condition-and-noise}
\begin{figure}[h!]
\centering
\includegraphics[width=0.9\textwidth]{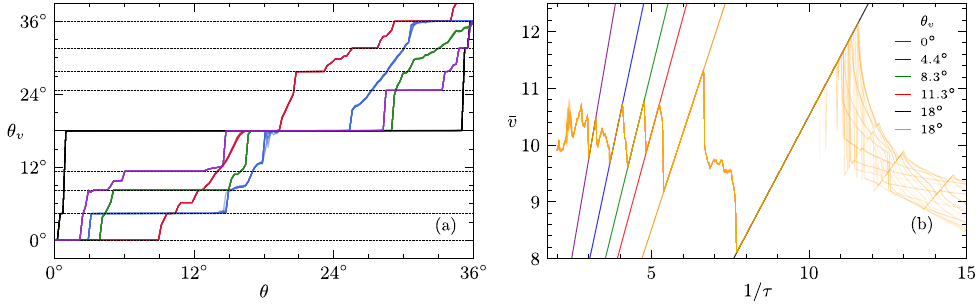}\vspace{-1ex} 
\caption{Directional and speed locking for 100 initial positions $\bm
  r(0)$ drawn randomly from the disk $|\bm r(0)|\le10$ and the same
  driving parameters as in Figs.~2(a) and 4(a).}
\label{fig:impact-initial-conditions}
\end{figure}

The data shown in the manuscript are for the initial condition $\bm
r(0)=\bm 0$, i.e.\ the particle starts its motion at the origin. Other
initial conditions lead to nearly identical directional locking and
also to nearly identical speed locking at driving frequencies
$1/\tau\lesssim10$.  This is demonstrated in
Fig.~\ref{fig:impact-initial-conditions}, where results are shown for
100 initial positions $\bm r(0)$ drawn randomly from the disk $|\bm
r(0)|\le10$. In Fig.~\ref{fig:impact-initial-conditions}(a), all 100
curves nearly overlap, i.e.\ the directional locking is not affected.
In Fig.~\ref{fig:impact-initial-conditions}(b), the curves are almost
identical for $1/\tau\lesssim10$. At higher driving frequencies, a
dependence of the mean particle speed on the initial condition occurs.

\begin{figure}[h!]
\centering
\includegraphics[width=0.9\textwidth]{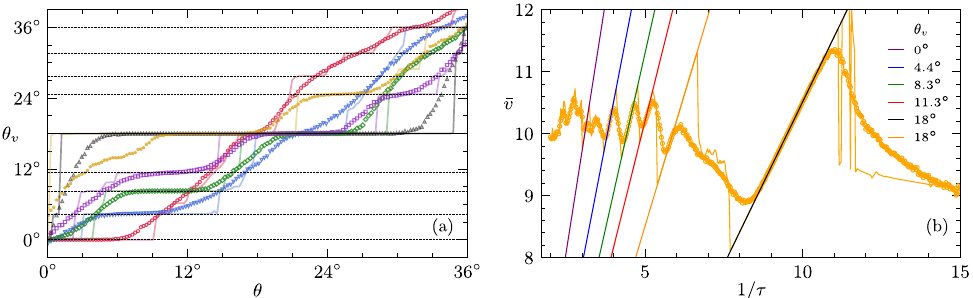}\vspace{-1ex} 
\caption{Directional and speed locking for noise (symbols) with
  strength $D=0.01$ [Eq.~(11)] in comparison with the zero-noise limit
  (thin lines). Parameters are the same as in Figs.~2(a) and 4(a) of
  the main text. In (a) we added the additional yellow curve, which is
  for the same parameters as the purple curve in Figs.~2(a) of the
  main text, but for the ellipse of the ac-driving rotated by
  $36^\circ$ in counterclockwise direction.}
\label{fig:impact-noise}
\end{figure}

Figure~\ref{fig:impact-noise} shows the impact of noise on the
directional and speed locking.  The noise decreases the extent of
locked regimes but most of the locked directions and speeds remain
clearly visible.

\vspace{2ex}

\vspace*{-2ex}



%

\end{document}